# Modeling of nucleobase/oligonucleotide interaction with graphene and graphene oxide: the role of charging and/or oxidizing the graphene surface

M.V. Karachevtsev[1], S.G. Stepanian[1], L. Adamowicz[2], and V.A. Karachevtsev[1]

[1]B.Verkin Institute for Low Temperature Physics and Engineering of the National Academy of Sciences of Ukraine

47 Nauky Ave., Kharkiv, 61103, Ukraine

[2]Department of Chemistry and Biochemistry, University of Arizona, Tucson, AZ 85721, USA

Corresponding authors:
Dr. Maksym V. Karachevtsev,
B.I. Verkin Institute for Low Temperature Physics and Engineering, of the National Academy of Sciences of Ukraine, 47, Nauky Ave., 61103 Kharkiv, Ukraine
Email: mkarachevtsev@ilt.kharkov.ua
Tel:+ (380)-57-340-22-23
Fax: + (380)-57-340-33-70

**Abstract**

We analyze the influence of the charge and the degree of oxidation of the surface of graphene (Gr) on its interaction with cytosine and oligonucleotide $r(C)_{10}$. This is a computational study involving DFT calculations and molecular dynamics simulations. It is shown that cytosine interacts stronger with graphene oxide (GO) than with Gr, while the energy of the interaction of cytosine with GO only weakly depends on the degree of the Gr oxidation. A correlation between the shifts of vibrational frequencies of cytosine due to complexation with GO and the degree of the Gr oxidation is found. The adsorption of anionic oligonucleotide $r(C)_{10}$ onto neutral and positively charged surfaces has a certain conformational similarity to conformations formed with similar van der Waals interactions. Also, for charged surfaces, the Coulombic attraction gives a noticeable contribution to the total interaction energy. For a negatively charged graphene surface the electrostatic repulsion between Gr surface and negatively charged phosphate backbone of oligonucleotide weakens the total binding energy. Competition between the Coulombic repulsion and the van der Waals attraction results in formation of a unique oligonucleotide conformation where all 10 cytosines are stacked onto Gr.

# 1. Introduction

In recent years, multiple studies have been devoted to biocompatibility of graphene and the family of graphene nanomaterials and to their possible applications in biology and medicine. Graphene oxide (GO) is a water-soluble derivative of graphene, that can be used as ideal scaffold for designing different nanobiostructures with possible applications in nanomedicine [1]. Biological and biomedical applications of these nanomaterials such as those in drug/gene delivery, cancer nanotherapeutics, bioimaging, tissue engineering, and bioelectronics have been reviewed in recent papers [2-4]. Among the applications, adsorption of DNA/RNA by GO is a particularly interesting due to its possible role in genosensing and gene delivery.

A few recent computational and experimental studies have carried out investigations of the interaction of DNA/RNA or its components with graphene [5-11]. The theoretical studies have mainly focused on determining the structures of the hybrids and the binding strengths of nucleobases with graphene (see, for example, ref 12 and the references cited therein). However, there have been smaller works devoted to studying the interaction of DNA and its components with graphene oxide because the task becomes more complex with the addition of the oxygen containing groups. In addition to the hydrophobic fragments of graphene, GO also possesses oxygen-containing functional groups such as hydroxyl, epoxy, carbonyl, and carboxyl groups that are attached at the plane or at the edges of the system. The more complex structure of GO makes the theoretical calculations more difficult, as it becomes necessary to consider multiple possible structures of GO and their numerous adducts with the biomolecules [8]. Thus understanding the interactions of biological molecules with GO is far from completion required significant computer resources to acquire such understanding.

Another aspect of the genosensing involving graphene nanomaterials is related to the unique properties of the carbon surface. For example, it is possible to exploit the graphene nanopores for DNA sequencing by employing translocation of the DNA strands through these nanopores [13, 14]. Under experimental conditions, the surface of such a nanosensor is usually charged and thus, in addition to the van der Waals interactions of graphene with biomolecules, electrostatic interactions also contribute to the total interaction energy. Thus, a complete account of the different effects contributing to the interaction of biopolymers with graphene is key to understanding this interaction and to apply biopolymer-graphene nanomaterials in genosensing.

We have already studied the interaction between a relatively long homopolynucleotide polyribocitydylic acid (poly(rC)) and GO by experimental methods and by theoretical calculations [8]. A spectral manifestation of the biopolymer binding with GO was obtained from the analysis of the UV absorption spectrum of the polymer bound to graphene oxide, as well as from the analysis of the FTIR spectrum of this complex. A change of the poly(rC) thermostability due to the adsorption on GO was also revealed in the study. Molecular dynamics simulations of the adsorption process of the $r(C)_{10}$ and $r(C)_{30}$ oligonucleotides on graphene demonstrated a disorder that occurs due to the π-π stacking of cytosines on graphene. The simulations also showed that longer oligonucleotides adsorb slower and that cytosine adsorbed to graphene oxide is additionally stabilized by H-bonding. In the graphene oxide model used in the calculations only two peripheral carboxylic groups (-COOH), two of the epoxy oxygen atoms (-O-) of the bases, and two hydroxyl groups (-OH) of the bases were considered. In the present work, as a next step, a more complicated GO model that includes 24 oxygen groups in its structure is used. The additional groups enhance the adsorption ability of cytosine. In addition to the analysis of the structures and the interaction energies, we also analyze the changes is the IR spectrum of cytosine due to adsorption to GO. The molecular dynamics simulation of the adsorption process of the $r(C)_{10}$ oligonucleotide onto a charged graphene surface shows a difference in the binding energy relative to the neutral surface. It also elucidates the influence of the charge on the structural transformation of oligonucleotide upon adsorption on graphene was shown as well. The poly(rC) as a model system has been often used before to study the DNA/RNA structural transformations occurring upon adsorption on different nanostructures including carbon nanomaterials [8]. Recently it was demonstrated that polynucleotide polyC has a much stronger affinity than other DNA homopolymers for nanocarbons [15]. High affinity of polyC to the GO surface can be used in biosensing and additional knowledge about polyC-GO hybrids is needed. Another reason for selecting this system as a model in the present study is also related to the possible use of GO as a scaffold for the delivery of the double-stranded poly(rI)·poly(rC) duplex which plays an important role in the biology and medicine as agent that activates the human innate immune system.

2. Methods

*Quantum-chemical calculations.* Structures, interaction energies, and vibrational spectra are calculated for the cytosine complexes with graphene oxide. In the calculations the B3LYP density-functional method [16-17] with an empirical dispersion correction is used. The D3 version of Grimme's dispersion with the Becke-Johnson damping [18, 19] invoked via the 'GD3BJ' keyword is employed. The standard 6-31++G(d,p) basis set is used in all calculations. The '5d' keyword is used to reduce the number of orbitals in each d-shell to five.

A hexagonal 96-atom fragment of the carbon surface with terminal hydrogen atoms is used in the calculations as a model of graphene. The models of graphene oxide are built by addition of oxygen containing groups (basal epoxy oxygen atoms (-O-) and basal hydroxyl groups (-OH)) to the graphene model. We use two graphene oxide models which differ in the number of the oxygen containing groups (4 or 24). The models are denoted as GO4 and GO24, respectively. In the models, the basal groups are placed on both sides of the graphene sheet. In GO24 the basal groups are randomly distributed over the whole graphene surface.

The interaction energies for the GO4-Cyt and GO24-Cyt complexes were calculated with the inclusion of the ZPVE and BSSE [20] corrections. All calculations were performed using the Gaussian 16 program package [21].

*Molecular Dynamics Simulation Protocol.* In the simulations of a graphene sheet containing 5682 C and 212 H atoms is used. All atoms of the graphene sheet lie in the same plane. Before starting the modeling, oligonucleotide $r(C)_{10}$ (in A-form) is placed near graphene and the distance between the graphene sheet and $r(C)_{10}$ is set to be more than 7Å. During the MD simulation the coordinates of the graphene atoms are frozen. The MD simulations are performed for three systems, which are different in terms of charges placed on the graphene atoms. In the first system, the charges on the graphene's carbons are -0.01e and on the hydrogens the charges are 0.26e (12 sodium ions are added to neutralize the overall charge of the system). In the second system the charges on the graphene's carbons are +0.01e and the charges of the hydrogens are -0.26e (8 sodium ions are added to neutralize the overall charge of the system). In the third system graphene's hydrogens and carbons are not charged (to neutralize the charge on the sugar-phosphate backbone 10 sodium ions are added). Each system is embedded in water (the number of water molecules varies from 28219 to 28223). The dimensions of the boxes for the simulation are 148Å×145Å×45Å. Program package NAMD [22] with Charmm force-field parameter set [23] is used in the simulations. Throughout the simulation, the number of atoms, the pressure, and the temperature of the system are kept constant (the NPT approach). During the simulation periodic boundary conditions are applied. In the periodic box the modeling temperature is 303 K and the pressure is 1 atm. To account for the electrostatic interaction, the Particle Mesh Ewald [24] method is used. The total simulation time is 10 ns with the simulation step equal to 1 fs. For visualization of the simulated nanostructures, software package VMD [25] is employed.

## 3. Results and discussion

### 3.1. Quantum-chemical calculations of the graphene oxide-cytosine complex

***Structures and interaction energies of graphene-oxide cytosine complexes.*** Fully optimized structures of GO4 and GO24 are shown in Fig. 1. As it is seen, addition of four oxygen-containing groups to the graphene fragment does not significantly change its planar structure. The deviations of the carbon atoms in GO4 from the planarity do not exceed 0.3 Å. The deviations are the largest for the carbon atoms bound to the oxygen-containing groups. For the remaining carbon atoms, the deviations are almost zero. Addition of 24 oxygen-containing groups (6 epoxy and 6 hydroxy groups on each side of the graphene surface) leads to a significant change of the structure of the carbon surface. This happens because the hybridization of carbon atoms changes from $sp^2$ to $sp^3$. As a result, there are practically no areas in GO24 where carbons form π-conjugated systems of bonds. This leads to an increased flexibility of the GO24 carbon backbone.

The most stable structures of the cytosine complexes with GO4 and GO24 are shown in Fig. 2. In the GO4-Cyt complex, the cytosine molecule is stacked on the pristine graphene surface and also forms two H-bonds with the oxygen containing groups. In the GO24-Cyt complex, only the H-bonding interaction between GO and cytosine is possible. The cytosine molecule forms multiple H-bonds due to presence of two hydrogen donor groups (the amino and imino groups) and two hydrogen acceptor atoms (the oxygen atom of the carbonyl group and the nitrogen atom). As it is seen in Fig. 2, all these groups are involved in H-bonding with oxygen-containing groups of GO24. It should be noted that for the GO24-Cyt system various structures with different H-bond networks are calculated. Their stabilities are very similar and only the most stable structure of the system is shown in Fig. 2. The interaction energies between cytosine and the models of graphene oxide are presented in Table 1. Data on the complex of cytosine with pristine graphene [8] are also included for comparison.

The interaction energies between the components of the GO4-Cyt and GO24-Cyt complexes are similar. Some additional calculations are performed to separate the contributions of the hydrogen bonds and the stacking interactions to the total interaction energy in the GO4-Cyt complex. In the calculations, the structure of GO4 is modified by removing oxygen-containing groups (but keeping the position of the cytosine with respect to graphene surface unchanged). With the removal of the oxygen-containing groups the calculated interaction energy is only due to the stacking interaction ($IE_{ST}$). The interaction energy due to the H-bonding ($IE_{HB}$) is estimated for each complex as the difference between the total interaction energy and the stacking interaction energy. The obtained data demonstrate that in the GO4-Cyt complex the main contribution to the interaction energy comes from the stacking interaction (-16.4 kcal/mol). At the same time, the absolute value of the energy of the stacking interaction is smaller than in the complex of cytosine with pristine graphene (-19.5 kcal/mol). This is due to the non-optimal arrangement of cytosine relative to the graphene surface in the former complex. In the GO24-Cyt complex, the interaction occurs only through the hydrogen bonding. In the GO24-Cyt complex, both proton donor and proton acceptor groups of cytosine form hydrogen bonds with numerous oxygen-containing groups of graphene oxide. As a result, the number of hydrogen bonds in GO24-Cyt is much larger than in GO4-Cyt (two H-bonds). This compensates for the absence of the contribution of the cytosine-graphene stacking interaction to the total interaction energy.

The similar interaction energies of cytosine with GO4 and GO24 demonstrate that, under experimental conditions, cytosine should be fairly evenly distributed over the surface of larger graphene oxide fragments even if they include areas with pristine graphene surface.

***Calculated IR spectra of the graphene oxide-cytosine complexes.*** Next, we performed a simulation of the IR spectra of the complexes found to determine the effect of the oxidation state of graphene on the spectral characteristics of the cytosine molecule and graphene oxide. Harmonic frequencies and IR intensities for the GO4-Cyt and GO24-Cyt complexes are calculated. Calculated IR spectra of the complexes as well spectra of cytosine and graphene oxide models are shown in Fig.3. The shifts of the cytosine frequencies in the fingerprint region calculated as the differences between the monomer frequencies and the corresponding frequencies of the cytosine moiety in the GO4-Cyt and GO24-Cyt complexes are presented in Table 2. The data shown in Fig. 3 demonstrate significant changes in the IR spectral characteristics of both cytosine and GO upon complexation. A general pattern is a decrease in the intensities of the vibrations of a cytosine molecule in the complexes compared to a non-complexed cytosine molecule. A similar intensity decrease was previously observed in complexes of cytosine with unoxidized graphene [8]. The effect was explained by the high polarizability of graphene associated with its π-electron system. As can be seen from Table 3, the decrease in the intensity of the cytosine vibrations in the complexes with GO is not as strong as in the complexes with unoxidized graphene. This can be explained by a partial compensation of the aforementioned intensity-decreasing effect by an increase of the vibrational intensities of those cytosine groups due to their involvement in hydrogen bonds with oxygen-containing graphene oxide groups.

Also, in the spectra of cytosine with highly oxidized graphene (GO24-Cyt), a pronounced increase in the vibrational intensities of oxygen-containing groups of GO is observed. This effect is particularly strong for the OH-bending and O-C-stretching vibrations of the hydroxyl groups (region 1500-1300 $cm^{-1}$). The effect is caused by the formation of hydrogen bonds of the C-OH groups with proton donor or proton acceptor groups of the cytosine molecule. A similar increase in intensity is observed for vibrations of the epoxy groups of GO24 in the region of about 1100 $cm^{-1}$. The shifts of the vibrational frequencies of cytosine caused by complexation show a sharp dependence on the degree of oxidation of graphene. Thus, the down-shift of the frequency of the C=O stretching vibration of cytosine is 18 $cm^{-1}$ in the complexes with pristine graphene. At the same time, in the complexes with GO4 and GO24, the shift becomes 48 and 78 $cm^{-1}$, respectively. In general, the strongest shifts in the vibrational frequencies are observed for those groups of cytosine and graphene oxide that are directly involved in the formation of intermolecular hydrogen bonds.

**3.2. Molecular dynamics modeling of r(C)$_{10}$ oligonucleotide adsorption on neutral and charged graphene**

In the simulations we use oligonucleotide containing 10 cytosines: r(C)$_{10}$. At the beginning of the simulation, the oligonucleotide in the self-ordered helical A-form is placed near the graphene surface (Fig. 4) and three cases of the oligonucleotide interaction with different graphene surfaces: neutral, positively charged and negatively charged are considered. During the simulation the conformation of the oligonucleotide (Fig. 5) and the interaction energy (Fig. 6) between graphene and the oligonucleotide are monitored.

The interaction of r(C)$_{10}$ with the carbon surface leads to the disruption of the self-ordered oligonucleotide chain because the binding energy of cytosine to the graphene surface is higher compared to the cytosine self-stacking. For each Gr surface, the oligonucleotide disordering proceeds in the simulation at a certain characteristic rate but with some peculiarities. For the neutral Gr surface the binding energy strongly increases after 5 ns simulation (up to -110 kcal/mol) as four cytosines become stacked with graphene (Fig. 6). At 20-25 ns the oligonucleotide reaches the most energetically favored conformation on the graphene surface with the binding energy of about -147 kcal/mol with seven cytosines stacked with graphene (Fig. 5b). For the last 10 ns of the simulation the oligonucleotide remains in this most stable conformation on graphene (Fig. 6). Thus one nucleotide interacts with the graphene surface with energy about -15 kcal/mol.

For positively charged graphene surface, due the electrostatic attraction with negatively charged phosphate backbone of the oligonucleotide, the interaction energy becomes stronger almost immediately after the simulation starts (Fig. 6). After a few nanoseconds the interaction energy becomes larger than -100 kcal/mol (Fig. 6). The destruction of the self-ordering of the oligonucleotide leads to its elongation and seven cytosines become stacked with graphene. The total binding energy of the oligonucleotide with the positively charged graphene surface is about -260-270 kcal/mol. The contributions from the van der Waals and Coulomb interactions to this energy are -148 kcal/mol and -123 kcal/mol, respectively.

For negatively charged graphene surface, its electrostatic repulsion with the negatively charged phosphate backbone of the oligonucleotide makes its achieving the most energetically favored conformation on graphene more difficult. Starting from a configuration with small, but positive, value of the interaction (repulsion) energy, some small attraction appears after 2-3 ns of the simulation. The attractive interaction energy increases during the whole simulation and reaches about -75 kcal/mol at 30 ns. This energy value is a sum of an attractive van der Waals interaction (-197 kcal/mol) and a repulsive Coulomb interaction (+123 kcal/mol). The stronger attractive interactions than the repulsion is a result of a unique conformation of the oligonucleotide stacked on the negatively charged graphene surface when all 10 cytosines are involved in the stacking (Fig. 5a). This configuration could be explained by the electrostatic repulsion lifting slightly the phosphate backbone over the surface of graphene and this uplift releasing some structural strains of oligonucleotide which appears due adsorption on the graphene surface. Also, note that 9 out of all 10 cytosines are arranged along one side of the backbone.

4. Conclusion

Calculations at the B3LYP(GD3BJ)/6-31++G(d,p) level of theory demonstrate a significant change in the structure of the carbon surface of highly oxidized graphene (with the O:C ratio of 24:96) due to oxidation. It is shown that cytosine interacts stronger with graphene oxide than with the pristine graphene. At the same time, the energy of interaction of cytosine with GO weakly depends on the degree of the graphene oxidation. Correlation between the shifts of the vibrational frequencies of cytosine due to complexation and the degree of the oxidation of graphene was found.

The influence of charging the graphene surface (with a positive or a negative charge) on the conformation of the $r(C)_{10}$ oligonucleotide adsorbed on the surface is compared with adsorption on a neutral surface. The interaction of $r(C)_{10}$ with the carbon surface leads to disruption of the oligonucleotide self-ordering and its replacement by cytosine stacking with graphene. Kinetics of the oligonucleotide adsorptions on the neutral and positive surfaces show certain similarity – in both cases at 30 ns of MD simulation seven cytosines become stacked with graphene. As a result, both oligonucleotide conformations have similar value of van der Waals interactions energy (147-148 kcal/mol), however, for the charged surface the Coulomb interactions also give a noticeable contribution (-123 kcal/mol) to the total interaction energy. For the negatively charged graphene surface, the electrostatic repulsion (of +123 kcal/mol) with negatively charged phosphate backbone of the oligonucleotide weakens the total binding energy (to about -75 kcal/mol). The competition between the Coulomb repulsion and the van der Waals attraction (-197 kcal/mol) results in a unique oligonucleotide conformation on the Gr surface with all 10 cytosines stacked with graphene.

The acquired knowledge on the conformations of oligonucleotide adsorbed on the graphene surface (charged and uncharged) and the corresponding binding energies can be useful for genosensing involved in scanning of the DNA sequences, as such scanning involves pulling biopolymer through the graphene pore with the assistance of an electrical potential. The information can also be useful in fabrication of biosensors based on graphene field-effect transistors with adsorbed DNA which constitutes the sensor recognition element.

**Acknowledgements**

This work was supported by the National Academy of Sciences of Ukraine (NASU) (Grant within the Joint research of National Academy of Sciences of Ukraine and Academy of Sciences of the Republic of Belarus "Nanobiostructures 2D nanomaterials with anticancer drugs: preparation and physical property study", Grant N0120U100157). M.V.K. acknowledges support from the NASU: Grant N 4/H-2019. The authors acknowledge the Computational Center at B. I. Verkin Institute for Low Temperature Physics and Engineering and High Throughput Computing (HTC) at the University of Arizona for providing computer time.

Table 1. BSSE and ZPVE corrected interaction energies (IE total, kcal/mol) of graphene-cytosine and graphene oxide-cytosine complexes. Decomposition of the interaction energies (only BSSE corrected) in the GO4-Cyt complex ($IE_{ST}$ and $IE_{HB}$, kcal/mol).[a]

| Complex | IE total | $IE_{ST}$ | $IE_{HB}$ |
|---|---|---|---|
| GR-Cyt [8] | -19.5 | | |
| GO4-Cyt | -27.7 | -16.4 | -10.9 |
| GO24-Cyt | -28.1 | | |

[a], all energies are calculated at the B3LYP(GD3BJ)/6-31++G(d,p) level of theory.

Table 2. Frequency shifts (Δν, cm$^{-1}$) in the fingerprint region (1800-500 cm$^{-1}$) calculated as difference between cytosine monomer frequency (ν, cm$^{-1}$) and frequency of cytosine moiety in the GR-Cyt, GO4-Cyt and GO24-Cyt complexes. [a]

| Monomer | | GR-Cyt [8] | | GO4-Cyt | | GO24-Cyt | | Assignment[b] |
|---|---|---|---|---|---|---|---|---|
| ν | I | Δν | I | Δν | I | Δν | I | |
| 1778 | 774.1 | -18 | 204.6 | -48 | 178.3 | -78 | 221.7 | C=O str |
| 1694 | 502.2 | -4 | 125.0 | -4 | 134.1 | -35 | 196.7 | C=C str |
| 1637 | 155.6 | +3 | 35.9 | +2 | 40.0 | -32 | 67.3 | NH$_2$ bend |
| 1573 | 170.1 | 0 | 42.8 | +6 | 25.6 | -27 | 14.3 | ring str, NH bend |
| 1511 | 140.2 | -2 | 49.1 | +1 | 55.0 | -29 | 79.2 | CH bend, NH$_2$ bend |
| 1446 | 80.9 | +2 | 22.3 | +20 | 17.2 | +106 | 41.5 | NH bend, ring str |
| 1360 | 47.4 | +1 | 15.8 | +8 | 13.8 | +8 | 80.1 | CH bend, NH bend |
| 1268 | 29.1 | +1 | 4.6 | +28 | 3.1 | +45 | 7.3 | NH bend, ring str |
| 1221 | 52.4 | +1 | 13.2 | +19 | 20.4 | +49 | 18.8 | NH bend, CH bend |
| 1129 | 3.2 | +6 | 0.3 | +8 | 0.1 | +6 | 0.4 | CH bend |
| 1090 | 41.0 | +18 | 13.3 | +31 | 9.2 | +52 | 12.0 | NH$_2$ bend, C=O bend |
| 990 | 0.4 | +7 | 0.2 | +12 | 0.6 | N/A[c] | N/A[c] | ring bend, NH$_2$ bend |
| 961 | 0.5 | +5 | 0.5 | +15 | 0.3 | N/A[c] | N/A[c] | CH tor |
| 927 | 3.9 | +11 | 0.6 | +28 | 1.0 | N/A[c] | N/A[c] | ring bend, NH$_2$ bend |
| 773 | 19.8 | +9 | 60.0 | +15 | 24.5 | +20 | 27.8 | CH tor |
| 770 | 36.6 | +5 | 24.8 | +13 | 49.6 | +21 | 45.4 | CH tor, ring tor |
| 765 | 2.9 | +6 | 8.1 | +13 | 4.5 | N/A[c] | N/A[c] | ring tor, CH tor |
| 724 | 31.1 | -1 | 20.7 | +21 | 52.9 | +35 | 60.6 | CH tor, NH tor |
| 625 | 66.0 | -2 | 103.1 | +120 | 64.6 | +115 | 71.1 | NH tor |
| 579 | 2.5 | +2 | 6.3 | +19 | 3.3 | N/A[c] | N/A[c] | ring bend |
| 546 | 2.8 | +4 | 10.1 | +7 | 4.4 | N/A[c] | N/A[c] | ring bend |
| 533 | 3.4 | +2 | 4.2 | +4 | 10.7 | N/A[c] | N/A[c] | ring bend, NH$_2$ tor |
| 525 | 11.4 | -3 | 65.9 | -2 | 59.4 | +24 | 69.3 | NH$_2$ tor |

[a] Frequencies and intensities (I, km/mol) are calculated at the B3LYP(GD3BJ)/6-31++G(d,p) level of theory. [b], str – stretching; bend – bending; tor – torsion. [c], This cytosine vibration is mixed with several vibrations of graphene oxide and cannot be unambiguously determined.

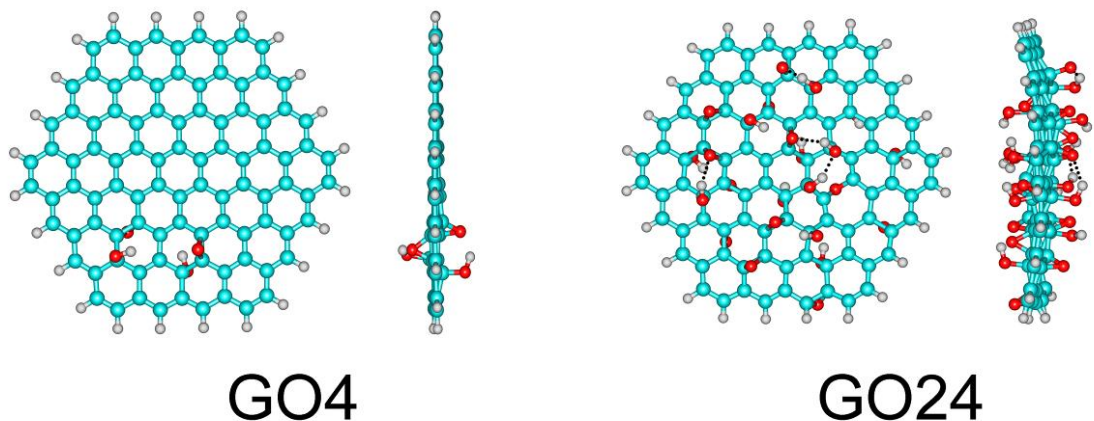

Figure 1. Structures (top view and side view) of the GO4 and GO24 models calculated at the B3LYP(GD3BJ)/6-31++G(d,p) level of theory.

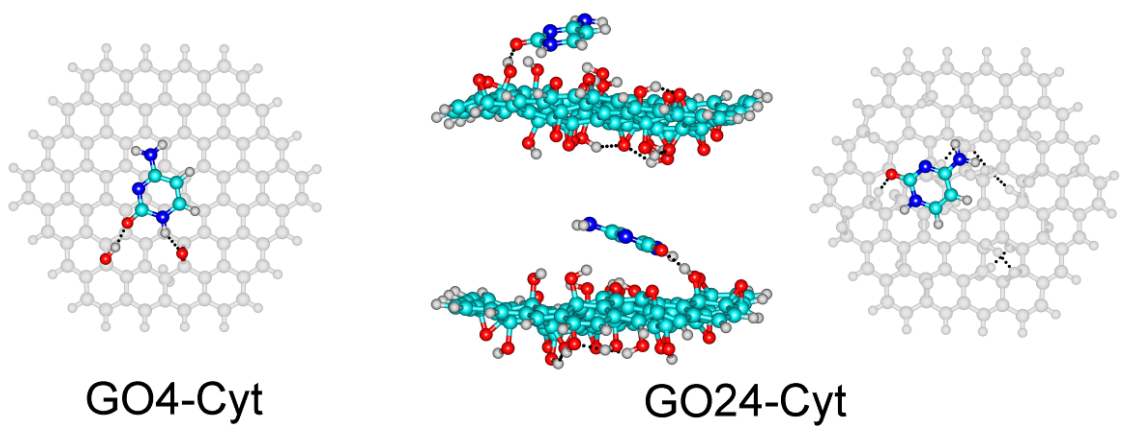

Figure 2. Structures of the GO4-Cyt (top view) and GO24-Cyt (two different side views and top view) complexes calculated at the B3LYP(GD3BJ)/6-31++G(d,p) level of theory.

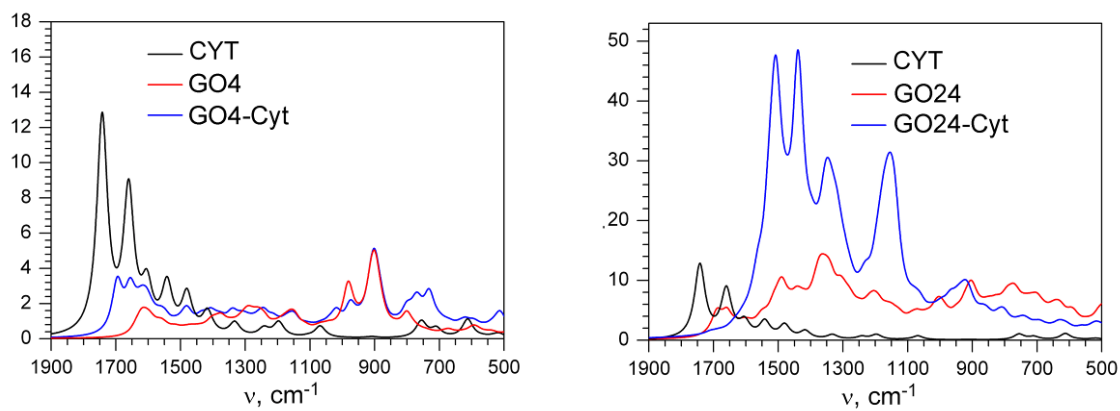

Figure 3. Graphical representations of the fingerprint regions of the calculated IR spectra of the GO4-Cyt and GO24-Cyt complexes. All frequencies are scaled down by 0.98. Bands are approximated by a 40 cm$^{-1}$ full-width-at-half-maximum Lorentzian lineshape. IR spectra are calculated at the B3LYP(GD3BJ)/6-31++G(d,p) level of theory.

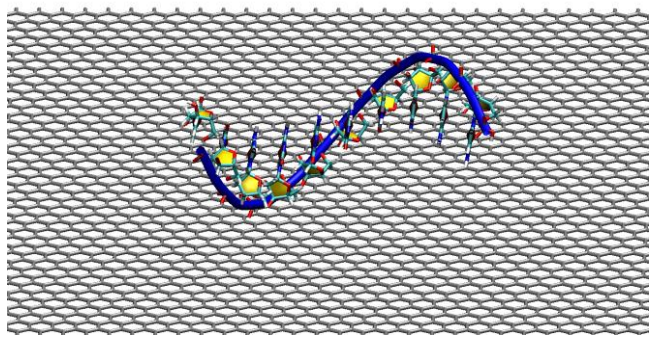

Figure 4. The starting location of the oligonucleotide r(C)$_{10}$ in the self-ordering helical A-form near the graphene surface.

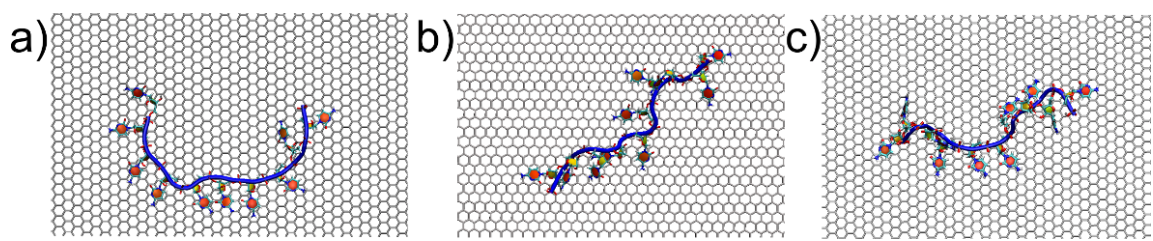

Figure 5. Snapshots of the hybrid formed by r(C)$_{10}$ adsorbed on graphene with the negative (a) neutral (b) and positive surface charges after 30 ns of the MD simulation.

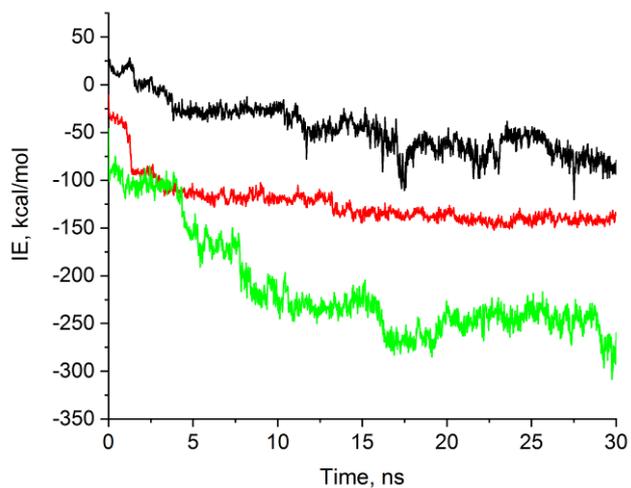

Figure 6. The interaction energy between r(C)$_{10}$ and graphene with the negative (black) neutral (red) and positive (green) surface charges as a function of the MD simulation time.